\documentstyle[preprint,aps]{revtex}

\begin{document}
\draft
\preprint{FNUSAL - 4/95}

\title{Does the quark cluster model predict any isospin
two dibaryon resonance?}

\author{A. Valcarce $^{(1)}$, H. Garcilazo $^{(2)}$,
F. Fern\'andez$^{(1)}$ and E. Moro$^{(1)}$}

\address{$(1)$ Grupo de F\' \i sica Nuclear \\
Universidad de Salamanca, E-37008 Salamanca, Spain}

\address{$(2)$ Escuela Superior de F\' \i sica y Matem\'aticas \\
Instituto Polit\'ecnico Nacional, Edificio 9,
07738 M\'exico D.F., Mexico}

\maketitle

\begin{abstract}
We analyze the possible existence
of a resonance in the $J^P=0^-$ channel with isospin
two by means of nucleon-$\Delta$ interactions based
on the constituent quark model. We solve the bound state and the
scattering problem using
two different potentials, a local and a non-local
one. The non-local potential results to be
the more attractive, although not enough
to generate the experimentally predicted resonance.

\end{abstract}

\narrowtext
\newpage

The existence of bound states of negative pions and
neutrons (pineuts) was predicted theoretically years
ago \cite{PIN}. A system of neutrons and negative pions
gives rise to a structure similar to an ordinary nucleus,
where the protons have been replaced by negative pions.
Since these systems can only decay through weak interactions,
they should be stable and have lifetimes comparable to that of
the pion. Then, a question arises immediately as to if one could
observe even the simple of these possible {\cal pineuts}, a
bound state of a pion and two neutrons, just like a nucleon-$\Delta$
bound state.

This problem has been studied from the theoretical point of
view by means of different methods \cite{SET,LAS} and with different
conclusions, but never definitively
excluding the possibility of a resonance.
Mainly due to this controversial
situation, a lot of experiments were done to find evidences
of such a resonance. After some signatures in experiments with
poor resolutions \cite{ASH}, an experiment with high intensity
proton beams almost excluded the existence of these structures
\cite{BOR}.

The situation has been recently renewed from the experimental point
of view. In Ref. \cite{BIL1}, a $J^P=0^-$ resonance
has been proposed to explain a sharp peak seen on the pionic
double charge exchange cross section in several nuclei from $^{14}C$
to $^{48}Ca$. The narrow width of these peaks suggested
that the resonance must have isospin even, otherwise
decay into nucleon-nucleon ($NN$) would be allowed, causing a much large
width. Besides, based on $QCD$ string models, they assumed
that the resonance has isospin zero.
However, in Refs. \cite{LAS,GAR1}
was pointed out that the narrow width of this structure could be related
with the vicinity of the nucleon-$\Delta$ ($N\Delta$) threshold and therefore
the resonance most likely must have isospin two (the $N\Delta$
system cannot coupled to isospin zero).

Our aim in this paper is to present the predictions of a quark
model based potential about the possibility of a $N\Delta$
resonance with isospin two. Previous calculations
based on the three-body formalism of the $\pi NN$ system
predict a large attraction in the
$0^-$ channel \cite{SET}, the same proposed
in Ref. \cite{BIL1}. Therefore,
the $0^-$ channel is the ideal candidate to
posses a resonance and we will concentrate on it.
Moreover, we will see that in case the resonance exists there is a
strong correlation between its mass and its
width due to the proximity of the $N\Delta$
threshold.

We have derived a $N\Delta$ interaction
using the same two-center quark-cluster model of Ref. \cite{VAL}.
Quarks acquire
a dynamical mass as a consequence of the chiral symmetry breaking.
To restore this symmetry one has at least to introduce
the exchange of a pseudoscalar (pion)
and a scalar (sigma) boson between quarks.
Besides, a perturbative contribution
is obtained from the non-relativistic reduction of
the one-gluon exchange diagram in QCD.

Therefore, the ingredients of the quark-quark
interaction are the confining potential (CON),
the one-gluon exchange (OGE), the
one-pion exchange (OPE) and the one-sigma exchange (OSE).
The explicit form of these
interactions is given by
(see Ref. \cite{VAL} for details),

\begin{eqnarray}
V_{CON} ({\vec r}_{ij}) & = &
-a_c \, {\vec \lambda}_i \cdot {\vec
\lambda}_j \, r^2_{ij} \,  , \\
V_{OGE} ({\vec r}_{ij}) & = &
{1 \over 4} \, \alpha_s \, {\vec
\lambda}_i \cdot {\vec \lambda}_j
\Biggl \lbrace {1 \over r_{ij}} -
{\pi \over m^2_q} \, \biggl [ 1 + {2 \over 3}
{\vec \sigma}_i \cdot {\vec
\sigma}_j \biggr ] \, \delta({\vec r}_{ij})
- {3 \over {4 m^2_q \, r^3_{ij}}}
\, S_{ij} \Biggr \rbrace \, , \\
V_{OPE} ({\vec r}_{ij}) & = & {1 \over 3}
\, \alpha_{ch} {\Lambda^2  \over \Lambda^2 -
m_\pi^2} \, m_\pi \, \Biggr\{ \left[ \,
Y (m_\pi \, r_{ij}) - { \Lambda^3
\over m_{\pi}^3} \, Y (\Lambda \,
r_{ij}) \right] {\vec \sigma}_i \cdot
{\vec \sigma}_j + \nonumber \\
 & & \left[ H( m_\pi \, r_{ij}) - {
\Lambda^3 \over m_\pi^3} \, H( \Lambda \,
r_{ij}) \right] S_{ij} \Biggr\} \,
{\vec \tau}_i \cdot {\vec \tau}_j \, , \\
V_{OSE} ({\vec r}_{ij}) & = & - \alpha_{ch} \,
{4 \, m_q^2 \over m_{\pi}^2}
{\Lambda^2 \over \Lambda^2 - m_{\sigma}^2}
\, m_{\sigma} \, \left[
Y (m_{\sigma} \, r_{ij})-
{\Lambda \over {m_{\sigma}}} \,
Y (\Lambda \, r_{ij}) \right] \, .
\end{eqnarray}

The main advantage of this model comes from the fact that
it works with a single $qq$-meson vertex. Therefore,
its parameters (coupling constants, cut-off masses,...) are
independent of the baryon to which the quarks are coupled, the
difference among them being generated by SU(2) scaling. This
makes the generalization of the $NN$ interaction to any other
non-strange baryonic system straightforward, and in particular
to the $N\Delta$ system.

Once the quark-quark interaction is chosen, an effective
nucleon-$\Delta$ potential can be obtained
as the expectation value of the energy of the
six-quark system minus the self-energies of the two clusters,
which can be computed as the energy of the six-quark system
when the two quark clusters do not interact:

\begin{equation}
V_{N \Delta (L \, S \, T) \rightarrow N \Delta (L' \, S' \, T)} (R,R') =
\xi_{L \,S \, T}^{L' \, S' \, T} (R,R') \, - \,
\xi_{L \,S \, T}^{L' \, S' \, T} (\infty,\infty) \, ,
\label{Poten1}
\end{equation}

\noindent
where,

\begin{equation}
\xi_{L \, S \, T}^{L' \, S' \, T} (R,R') \, = \,
{{\left \langle \Psi_{N \Delta}^{L' \, S' \, T} ({\vec R}') \mid
\sum_{i<j=1}^{6} V_{qq}({\vec r}_{ij})
\mid \Psi_{N \Delta}^{L \, S \, T} ({\vec R}) \right \rangle}
\over
{\sqrt{\left \langle \Psi_{N \Delta}^{L' \, S' \, T} ({\vec R}') \mid
\Psi_{N \Delta}^{L' \, S' \, T} ({\vec R}') \right \rangle}
\sqrt{\left \langle \Psi_{N \Delta}^{L \, S \, T} ({\vec R}) \mid
\Psi_{N \Delta}^{L \, S \, T} ({\vec R}) \right \rangle}}} \, .
\label{Poten2}
\end{equation}

\noindent
The parameters of the model are those of Ref. \cite{VAL}. As local
potential we will assume $R=R'$.

In order to determine the nature (attractive or repulsive)
of the $0^-$ $N\Delta$ channel,
we will first calculate the Fredholm determinant of that channel
as a function of energy assuming a stable
delta and nonrelativistic kinematics. That means, we will use the
Lippmann-Schwinger equation

\begin{equation}
T_{ij}(q,q_0)=V_{ij}(q,q_0)+\sum_k \int_0^\infty q^{\prime 2} dq'
V_{ik}(q,q')G_0(E,q')T_{kj}(q',q_0) \, ,
\end{equation}

\noindent
where the two-body propagator is

\begin{equation}
G_0(E,q)={1 \over E-q^2/2\eta+i\epsilon} \, ,
\label{prop}
\end{equation}

\noindent
with reduced mass

\begin{equation}
\eta={m_Nm_\Delta \over m_N+m_\Delta} \, .
\end{equation}

\noindent

The energy and on-shell momentum are related as

\begin{equation}
E=q_0^2/2\eta \, ,
\end{equation}

\noindent
and we will restrict ourselves to the region $E \le 0$.

If we replace the integration in Eq. (7) by a numerical quadrature, the
integral equations take the form

\begin{equation}
T_{ij}(q_n,q_0)=V_{ij}(q_n,q_0)+\sum_k \sum_m w_m q^2_m
V_{ik}(q_n,q_m)G_0(E,q_m)T_{kj}(q_m,q_0) \, ,
\end{equation}

\noindent
where $q_m$ and $w_m$ are the abscissas
and weights of the quadrature (we
use a 40-point Gauss quadrature).
Eq. (11) gives rise to the
set of inhomogeneous linear equations

\begin{equation}
\sum_k \sum_m M^{ik}_{nm}(E)T_{kj}(q_m,q_0)=V_{ij}(q_n,q_0) \, ,
\end{equation}
with

\begin{equation}
M_{nm}^{ik}(E)=\delta_{ik}\delta_{nm}- w_m q^2_m
V_{ik}(q_n,q_m)G_0(E,q_m) \, .
\end{equation}
If a bound state exists at an energy $E_B$,
the determinant of the matrix
$M_{nm}^{ik}(E_B)$ (the Fredholm determinant)
must vanish, i.e.,

\begin{equation}
\left|M_{nm}^{ik}(E_B)\right|=0.
\end{equation}

Even if there is no bound state, the Fredholm
determinant is a very useful
tool to determine the nature of a given channel.
If the Fredholm determinant is
larger than one that means that channel is repulsive. If the Fredholm
determinant is less than one that means the channel is attractive.
Finally, if the Fredholm determinant passes through zero that means
there is a bound state at that energy.

In Figure 1 we compare the Fredholm determinant generated
by the local and non-local quark model based potentials.
The non-locality of the interaction
generates additional attraction, enough to produce
a resonance (it goes through zero).

To determine the exact location of the resonance,
we calculate Argand diagrams
between a stable and an unstable particle
using the formalism
of Ref. \cite{GART}.
In this case, however, we will use relativistic kinematics
and will include the width of the delta. That means, instead of the
propagator (\ref{prop}) we will use \cite{GART}

\begin{equation}
G_0(S,q)={2m_\Delta \over s-m_\Delta^2+im_\Delta\Gamma_\Delta(s,q)} \, ,
\end{equation}

\noindent
where $S$ is the invariant mass squared of the system,
while $s$ is the
invariant mass squared of the $\pi N$ subsystem (those
are the decay products
of the $\Delta$) and is given by

\begin{equation}
s=S+m_N^2-2\sqrt{S(m_N^2+q^2)} \, .
\end{equation}

The width of the $\Delta$ is taken to be \cite{GART}

\begin{equation}
\Gamma_\Delta(s,q)={2 \over 3}\,0.35\, p_0^3 {\sqrt{m_N^2+q^2}\over
m_\pi^2\sqrt{s}} \, ,
\end{equation}

\noindent
where $p_0$ is the pion-nucleon relative momentum given by

\begin{equation}
p_0=\left([s-(m_N+m_\pi)^2][s-(m_N-m_\pi)^2] \over 4s \right)^{1/2} \, .
\end{equation}

We show in Figure 2 the phase shifts for the
$0^-$ channel. As it can be seen from this figure, the
attraction is only strong enough to produce a
resonance with the non-local potential (it reaches 90 degrees).
This resonance lies at 2145.6 MeV and has a width
of 148.12 MeV for the
mass of the sigma predicted
by chiral symmetry requirements $m_\sigma \sim 675$ MeV.

The proposed resonance has a mass of 2065 MeV
and a very small width of
0.51 MeV \cite{BIL1}. It is therefore very
interesting to investigate whether the
 nucleon-$\Delta$ system exhibit
the features of this resonance, and
particularly such a tiny width.

In order to do this, we have artificially varied the mass of the
$\sigma$ meson with both potential models,
such as to increase the amount of attraction. We
show in Table I the mass and width of the resonance
and the corresponding mass of the sigma meson necessary to generate
it. The width of the
resonance drops dramatically when its mass approaches the
$\pi NN$ threshold (2017 MeV).
This result can be understood from simple angular momentum barrier
considerations. If we call $q$ and $L$ to the relative momentum and
relative orbital angular momentum between a nucleon and the $\pi$-nucleon
pair, respectively, then since $L=1$ the
width of the resonance will be proportional
to $q^{2L+1}=q^3$, so that it will drop very fast as one approaches the
$\pi NN$ threshold since there $q \to 0$.

In both local and non local potential models, when the mass of the sigma is
taken to reproduce the predicted mass of the resonance
(2065 MeV) the width is very narrow,
which is in very good agreement with
the predictions extracted by Bilger
and Clement \cite{BIL1}.
Therefore, the sharp peak seen in the
double charge exchange reactions could be justified as a
nucleon-$\Delta$ resonance in the isospin 2 channel, without
resorting to other more exotic processes.

As a summary, we have studied the
nucleon-$\Delta$ system in the $0^-$ channel with isospin two,
within the quark cluster model of the baryon-baryon interaction. We
have used a local and a non-local potential.
We found that the non-local effects generate
additional attraction, although not enough
to reproduce the resonance
predicted in Ref. \cite{BIL1}.
However, due to the proximity of the
nucleon-$\Delta$ threshold, if we force the resonance mass to
reach the experimental predicted value of 2065 MeV,
then its width is very narrow, in very good
agreement with the width extracted in Ref. \cite{BIL1}.

\acknowledgements

This work has been partially funded by
EU project ERBCHBICT941800,
DGICYT Contract No. PB91-0119 and by
COFAA-IPN (Mexico).

\begin{table}
\caption{ Mass and width of the $0^-$ resonance with the
corresponding mass of the sigma meson using the local
and non-local nucleon-$\Delta$ potentials based on the
constituent quark model.}
\label{width}

\begin{tabular}{ccccc}
{\rm Potential model} & $m_\sigma (MeV)$       & $M_{Res} (MeV)$
&  $\Gamma_{Res} (MeV)$& \\
\tableline
{\rm Local} & 234.0                            &  2064.4
&  0.6           & \\
{\rm Non local} & 422.0                         &  2064.5
&  1.6           & \\
\end{tabular}
\end{table}

\end{document}